\documentclass[twocolumn,showpacs,showkeys,preprintnumbers,amsmah,amssymb]{revtex4}
\usepackage{graphicx}
\usepackage{dcolumn}
\usepackage{bm}

\begin{document}                

\title{Spin-resolved off-specular neutron scattering from magnetic domain walls
 using the polarized  $^3He$ gas spin filter}
\author{F.\ Radu$^{a}$$^,$%
\footnote{Permanent address:
Department of Experimental Physics,
National Institute of Physics and Nuclear Engineering,
P.\ O.\ BOX MG-6, 76900, Magurele, Romania.}$^,$%
\footnote{Corresponding author. Tel: +49-234-322-4835; Fax: +49-234-321-4173\\
Email address: {\tt Florin.Radu@rub.de} (F.\ Radu)}
A. Vorobiev$^{b}$, J. Major$^b$, H. Humblot$^{ c}$,  K. Westerholt$^a$, H. Zabel$^a$}
\affiliation{ ${^a}$Institut f\"{u}r Experimentalphysik/Festk\"{o}rperphysik,
Ruhr-Universit\"{a}t Bochum, 44780 Bochum,  Germany\\ $^b$Max-Planck-Institut   f\"ur Metallforschung, Heisenbergstr. 3, D-70569 Stuttgart, Germany \\${^c}$Institut Laue-Langevin, 38042 Grenoble Cedex 9, France}

%
\begin{abstract}

We report on the use of the polarized $^3$He gas filter and
neutron resonant enhancement techniques for the measurement of
spin-polarized diffuse neutron scattering due to ferromagnetic
domains. A CoO/Co exchange biased bilayer was grown on a
Ti/Cu/$Al_2O_3$  neutron resonator template. The system is cooled
in an applied magnetic field of $H_a=2000$~Oe through the N\'{e}el
temperature of the antiferromagnet to 10 K where the applied
magnetic field is swept as to measure the magnetic hysteresis
loop. After the second magnetization reversal at the coercive
field $H_{c2}=+ 230$~Oe, the system is supposed to approach the
original magnetic configuration. In order to prove that this is
not the case for our exchange biased bilayer, we have measured
four off-specular maps I$++$,  I$+-$,  I$-+$, I$--$  at  $H_a
\approx + 370$~Oe, where the Co magnetic spins were mostly
reversed. They show a striking behavior in the total reflection
region: while the nonspin-flip scattering exhibits no diffuse
reflectivity, the spin-flip scattering shows strong diffuse
scattering at incident angles which satisfy the resonance
conditions. Moreover the spin-flip off-specular part of the
reflectivity
 is asymmetric. The I$-+$ intensity occurs at higher exit angles than the
specularly reflected neutrons, and the I$+-$
intensity is shifted to lower angles.
Their intensities are  noticeably different and there is a
splitting of the resonance positions for the up and down neutron
spins ($\alpha_{n} ^{+} \ne \alpha_{n} ^{-}$) . Additionally, a
strong influence of the  stray fields from magnetic domains to the
resonance splitting is observed.\\



\end{abstract}


\maketitle


The exchange bias (EB) phenomenon is associated with an exchange coupling
between a ferromagnetic (F) and an antiferromagnetic (AF) layer across their
common interface, resulting in an unidirectional
 magnetic anisotropy and a shift of the magnetic hysteresis by an
exchange bias field $H_{EB}$~ \cite{nogu}. A remarkable effect related to the
exchange bias is the asymmetry of the magnetization reversal,
which is best revealed by neutron scattering~\cite{fitz,gier,lee,radu1}.
The first magnetization reversal after field cooling is dominated
 by domain wall movement, whereas all subsequent reversals proceed
essentially by rotation of the magnetization.  In the present work we
report on neutron off-specular diffuse scattering  experiments
performed at applied magnetic fields close to the
second coercive field $H_{ c2}$ supplementing the previous
results at fields near the
first coercive field $H_{ c1}$~\cite{radu1}.

The polarized $^3$He gas spin filter was recently used for the first
time~\cite{nick} in a reflectivity experiment to measure
diffuse scattering from thin magnetic films. The device itself is a cylindrical
cell of 100 mm in length and 50 mm in diameter, filled with 0.5~bar
$^3$He gas. After filling, the neutron polarization $P_n=P_p  P_{^3He}$ (where the $P_n$
is the measured neutron polarization, $P_p$ is the polarization of the polarizer and the $P_{^3He}$
 is the polarization of the analyzer)  was 65~\% and
it decayed to about 59 \% in 24 hours. The transmission of the
neutrons through the filter cell~\cite{fran} is spin dependent with the
following  cross-sections: $ \sigma_{\uparrow \uparrow} $=5~bn and
$ \sigma_{\uparrow \downarrow}$=16500~bn. In order to record the
scattered intensities corresponding to all four neutron scattering
cross-sections (I$++$, I$+-$, I$-+$, I$--$), the experimental
set-up includes, in addition, two Mezei spin flippers placed
upstream and downstream of the sample~\cite{nick}. The advantage
of using the $^3$He spin analyzer is that it is possible to
register the spin-resolved diffuse scattering for a broad range of
exit angles, and simultaneously it is free of additional spin-flip
or small angle scattering, as it may occur for broad angle
supermirror analyzers.  However, there are drawbacks as well,
namely that the cell absorption is not negligible and the flipping
ratio seemed to be lower as compared to the case of the
supermirror. In the present experiment these difficulties,
especially the reduced neutron transmission, were successfully
compensated by using an enhancement technique provided by the
sample design as a neutron resonator~\cite{ri,ir}.

Recently (Nov-2002), the neutron polarization of the $^3$He
spin-filter reached an outstanding value of 94\%~\cite{rhe3} with
a neutron transmission of about 40\%. Therefore the spin-flip
ratio became comparable to the one provided by the supermirror.
The experiment we report here was, however, performed on Sept-2001
when the $^3$He did not have such high value of polarization.

The neutron resonances in thin films can be generated in two different ways:
one way is in the total reflection region of a neutron resonator (bound states) and the other
 way is in a single film just above its critical angle for total reflection
  (quasi bound states),  i. e. in the region of
the Kissing fringes.
In both cases the neutron density in the system is enhanced~\cite{ri,ir} for definite
 incident neutron energies and one uses this phenomenon to increase the
sensitivity to magnetic and nonmagnetic inhomogeneities and to
non-collinear  magnetic spin arrangements at the interface. As
long as the observed effect takes place in the total reflection
region, the scattered intensity is essentially the same as the
incident intensity.

In the present experiments we used the neutron resonator to study the scattering
from ferromagnetic domains in a CoO(25\AA)/Co(200\AA) exchange biased bilayer.
The bilayer was grown on a Ti(2000~\AA)/Cu(1000~\AA)/Al$_2$O$_3$  neutron resonator template.
The enhancement of the neutron
density takes place in the
Ti ($\rho_{\rm Ti}=-1.946 \cdot 10^{-6}$~{\AA}$^{-2}$) layer. The
neutron scattering-length density $\rho$ is defined as $\rho=N b_{ coh}$,
$N$ is the density of the scattering nuclei and
$b_{ coh}$ is its coherent scattering length.
The tail of the resonance enhanced neutron wave function
extends over the CoO/Co interface providing an increased sensitivity to
its  magnetic properties.
An important contribution to the increased sensitivity
is the total-reflection region itself, provided by the
Cu ($\rho_{\rm Cu}=6.55 \cdot 10^{-6}$~{\AA}$^{-2}$) reflector layer.

\begin{figure}[th]
 \begin{center}

\includegraphics[clip=true,keepaspectratio=true,width=1\linewidth]{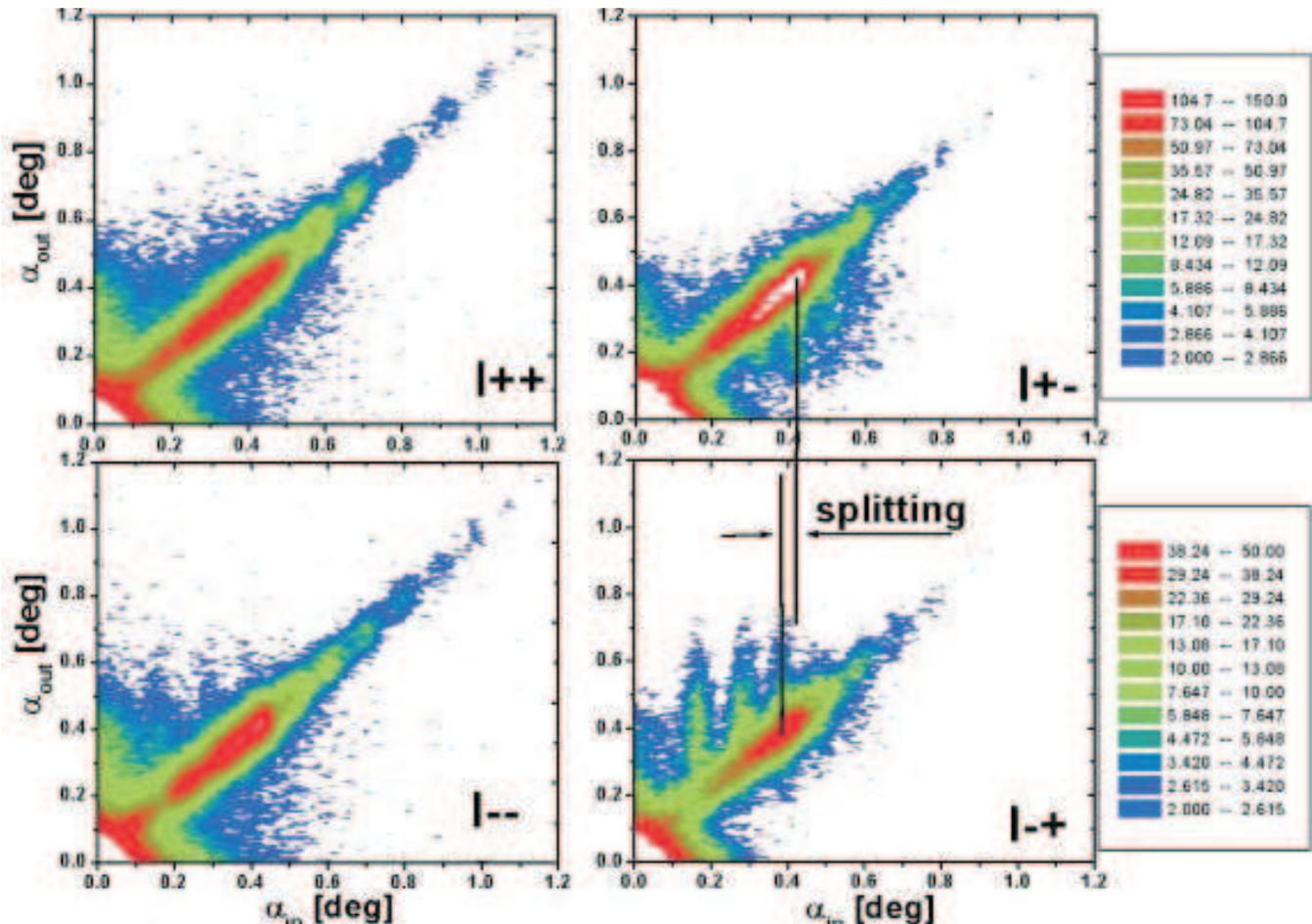}
\end{center}
\vspace{-1.0cm}
 \caption{\label{psdmaps} Off-specular scattering maps from CoO/Co/Ti/Cu/Al$_2$O$_3$. The sample
was cooled in an in-plane applied magnetic field of $H_a^{FC}=+ 2000$~Oe.
 At T=10 K the hysteresis loop was measured and the sample was set in a field of $H_a^=+ 370$~Oe,
 which is
 slightly higher  than  the second coercive field $(H_{c2}=+ 230$~Oe. The data it is not corrected
for spin-flip efficiencies.}
\end{figure}

The system is cooled in an in-plane applied magnetic field of
$H_a^{FC}=+ 2000$~Oe through the N\'{e}el temperature of the ferromagnet to
10 K, where the applied magnetic field $H_a$ is swept as to
measure the magnetic hysteresis loop. After the second
magnetization reversal ($H_{c2}=+ 230$~Oe), the system is supposed
to approach the original magnetic configuration. In order to prove
that
 this is not the case for our exchange biased bilayer,  we have measured four
off-specular neutron reflectivity maps   I$++$,  I$+-$,  I$-+$, I$--$  [Fig.~\ref{psdmaps}]
at  $H_a \approx + 370$~Oe, where the magnetic
 spins were mostly reversed back to the direction of the applied field.
 In Ref. \cite{radu1} the off-specular
 scattering (rocking curves at $\alpha_{\rm in} + \alpha_{\rm out}=$constant)
in the I$+ -$ map was
used to follow the development of the  magnetic  domains at the CoO/Co
interface before and after the first magnetization reversal to
 saturation and back to the remanence.
 In the present report we show measurements of the   magnetic state of the EB system
close but after the  second magnetization reversal.  The
experiments were performed at the EVA reflectometer of the
Institut Laue-Langevin (Grenoble, France) \cite{EVAref} with a
neutron wavelength of 5.5~{\AA} using a 1D position sensitive detector
(PSD).

\onecolumngrid@push
\vspace{0mm}

\twocolumngrid
The polarized neutron scattering scattering maps (Fig.
\ref{psdmaps})
 show a striking behavior in the total reflection region:
 while the nonspin-flip scattering shows weak diffuse reflectivity ,
 the spin-flip scattering exhibits
strong diffuse scattering at incident angles which satisfy the
 resonance conditions~\cite{ri,ir}. Each resonance in such a system  has
different enhancement factors, decreasing from low  orders (low $\alpha_{in}$) to
higher order (high $\alpha_{in}$). One can easily observe in the I$-+$ map that the off-specular
spikes are stronger in intensity  for a large enhancement factor~\cite{ri}.
The I$+-$ map shows a weaker intensity in the  off-specular scattering
signal, which is an effect of the
different optical potential for (+) and (-) states, where the (+) state
 ($\rho^+_{Co}=  6.3 \cdot 10^{-6}$~{\AA}$^{-2}$) denotes the parallel
orientation of the Co spins to the polarization direction and (-) state
 ($\rho^-_{\rm Co}=-1.78 \cdot 10^{-6}$~{\AA}$^{-2}$) denotes
the antiparallel alignment of the Co spins in respect to the
polarization direction.
 The enhancement factor for a (-) state is smaller
than for (+) state. This is caused by the transmission coefficient from vacuum
 through the top layers
 (CoO and Co) to the resonant layer (Ti). For
the (-) states the transmission coefficient is higher than for the (+) state and
this affects strongly the enhancement factor for the (+) and (-) state resonances
in the Ti layer.

The off-specular scattering is caused by the refraction on domain
walls~\cite{rhe3}. Assuming the domain walls to be perpendicular
to the sample surface, the spikes will disappear and the
off-specular intensity would be concentrated at their respective
ends distinctly separated from the specular ridge~\cite{rhe3}. In
order to explain a continuous spike we have to assume that the
domains walls are continuously distributed. Furthermore, the
orientation of the domain walls must spread from perpendicular to
parallel to the sample surface.  In a first approximation, the
domain walls could be considered as the magnetic roughness at the
CoO/Co interface.


Another striking effect in Fig.\ \ref{psdmaps} is the splitting of the position
 ($\alpha_{in}$) of the resonances by $0.68$ mrad ($=.039^\circ$) .
 The magnitude of the splitting is too strong
 to be caused by the small  applied field of $H_a=370$~Oe.
 The observed splitting requires an effective
 field of $1100$~Oe [see Ref~\cite{felcher,korneev}].
On the other hand, under the same experimental conditions of
temperature and field, and before the first reversal takes place,
i.e. no magnetic domains are created so far, we observe only a
small splitting, which is consistent with an applied field of
370~Oe (not shown). Therefore, we need to assume that an
additional stray field acts on the incoming neutrons and
consequently on those which are inside of the spacer layer (Ti).
With this assumption the value of the splitting becomes feasible.
The stray fields will not cause a spin-flip signal as they feature
a slow gradient, but they will change the incoming and exit
energies for the spin-flipped neutrons. Thus, we suggest that the
stray fields from the magnetic domains are adding up to the
applied field and the resulting Zeeman effect is the reason for
the splitting we observed.

The additional field generated by the stray fields and the
distribution of domain walls we observed leads us to the
assumption that the interfacial  magnetic domains are
ferromagnetic and that their magnetization vector is oriented
perpendicular to the sample plane. The other possible case when
the orientation of the magnetic domains were in-plane and
perpendicular to the applied field would not cause the required
stray fields for the splitting nor the type of domain walls
necessary for generating the  above described off-specular
intensity distribution (spikes).

 In conclusion we have used a polarized $^3$He gas filter and neutron resonant
 enhancement techniques for the measurement of spin-polarized
 diffuse neutron scattering from ferromagnetic domain walls of an exchange
 biased CoO/Co bilayer. After
 the first reversal domains are created at the CoO/Co interface,
 which remain further on. We have shown that the magnetization of those domains
 points perpendicular to the interface and that their domain wall(s) is continuously distributed
 from perpendicular to parallel  to the interface.
 We believe that  the enhancement of  off-specular scattering by using neutron resonators
opens a new direction towards small angle neutron scattering
(SANS)  studies of magnetic and nonmagnetic thin films. The
sensitivity to clusters, porous media, interface spin
misalignment,
 magnetic domain walls, vortices, and inhomogeneities is under these
 conditions remarkably high.

This work is supported by the Deutsche Forschungsgemeinschaft (DFG), SFB 491.

\end{document}